\documentclass[aps,pre,amsmath,amssymb,twocolumn,groupedaddress,showpacs]{revtex4}
\usepackage{graphicx}
\usepackage{amssymb}
\usepackage{amsmath}
\usepackage{epstopdf}
\newcommand{\be}{\begin{equation}}\newcommand{\ee}{\end{equation}}
\newcommand{\ba}{\begin{array}{l}}\newcommand{\ea}{\end{array}}
\newcommand{\baa}{\begin{eqnarray}}\newcommand{\eaa}{\end{eqnarray}}
\newcommand{\lab}[1]{\label{#1}}\newcommand{\re}[1]{(\ref{#1})}
\newcommand{\ci}[1]{\cite{#1}}

\begin{document}
\title{PT-symmetric quantum graphs}
\author{D. U. Matrasulov$^a$, K. K. Sabirov$^b$,  J. R. Yusupov$^a$}
\affiliation{$^a$ Turin Polytechnic University in Tashkent, 17
Niyazov Str., 100095,  Tashkent, Uzbekistan\\ $^b$Tashkent
University of  Information Technology, Amir Temur Avenue 108,
Tashkent 100200, Uzbekistan}

\begin{abstract}
We consider PT-symmetrically branched quantum wires, in which the
branching points  provide PT-symmetric boundary conditions for the
Schr\"odinger equation on a graph. For such PT-symmetric quantum
graph we derive general boundary conditions which keep the
Hamiltonian as PT-symmetric with real eigenvalues and positively
defined norm. Explicit boundary conditions which are consistent
with the general PT-symmetric boundary conditions are presented.
Secular equations for finding the eigenvalues of the quantum graph
are derived. Breaking of the Kirchhoff rule at the branching point
is shown. Experimental realization of PT-symmetric quantum graphs
on branched optical waveguides is discussed.
\end{abstract}

\maketitle

{\it Introduction.} PT-symmetric quantum systems attracted much
attention since from the pioneering paper \ci{CMB1}, where the
authors showed that a quantum system with non-Hermitian but
PT-symmetric Hamiltonian can have a set of eigenstates with real
eigenvalues (a real spectrum). Later it was strictly shown that
the Hermiticity of the Hamiltonian is not a necessary condition
for the realness of its eigenvalues. Quantum mechanics of such
systems has become rapidly developing topic by now and called
PT-symmetric quantum mechanics (see papers \ci{CMB2}-\ci{ptbox}
for review of recent developments on the topic). Different aspects
of PT-symmetric quantum physics have been studied in huge number
of papers published during past two decades. These studies allowed
to construct complete theory of PT-symmetric quantum system,
including PT-symmetric  field theory \ci{CMB07,CMB11}.
Experimental realization of such systems was also subject for
extensive research. The latter has been done mainly in optics
\ci{Makris,Nature,UFN,Konotop}. Some other PT-symmetric systems
are discussed recently in the literature \ci{Chit,Longhi}.
PT-symmetric relativistic system are also studied in
\ci{PTSD1,PTSD2}. General condition for PT-symmetry has been
derived in terms of so-called PT-symmetric inner product. However,
since such condition does not provide positively defined norm of
the eigenvalues,  its extension in terms CPT-symmetric inner
product was proposed in \ci{CMB5,CMB9,CMB11}. Similarly to the
case of Hermiticity, PT-symmetry can be introduced either through
the complex potential or by imposing proper boundary conditions
which provides such symmetry via the inner product
\ci{CMB5,CMB11}. Different types of complex potentials providing
PT-symmetry in Hamiltonian have been considered in
\ci{CMB9,CMB11}. Introducing PT-symmetry in terms of proper
boundary conditions was studied for particle-in-box system in
\ci{CMB13,Znojil,Znojil2,ptbox}. Certain progress is also done in
nonlinear extension of PT-symmetric systems
\ci{UFN,Konotop,Panos}. Spectral properties of the Laplace
operators on graph in the presence of PT- and reflection symmetry
have been considered in \ci{Kurasov1, Kurasov2}.

In this Letter we consider the problem of PT-symmetric quantum
graph which implies connecting quantum wires according to
PT-symmetry. The latter means imposing PT-symmetric boundary
conditions at the branching point of the quantum graph. Quantum
graph itself is a branched system of quantum wires. Branching
(connection) rule is called topology of a graph and given in terms
of adjacency matrix \ci{Uzy1,Uzy2}. When length are assigned to
the bonds of a graph, it is called a metric graph. The vertex
(branching)boundary condition for Schr\"odinger equation on a
quantum graph are imposed as providing Hermiticity of the
Hamiltonian. First strict study of quantum graphs as branched
quantum wires was presented in \ci{Exner1}. General boundary
conditions for the Schr\"odinger equation on graphs were derived
in terms of Hermitian inner product \ci{Kost}. Later such boundary
condition have been derived for Dirac equation on graphs in
\ci{Bolte}. Spectral statistics and manifestation of quantum chaos
was studied in \ci{Uzy1,Uzy2}. Different aspects of the
Schr\"odinger operator on graphs have been studied in the
Refs.\ci{Kuchment04}-\ci{Exner15}. Experimental realization of
quantum graphs on optical microwave networks has been discussed in
\ci{Hul}. Nonlinear extension of the wave dynamics in networks
considered in \ci{Our1,Our2,Our3}. Here we derive PT- symmetric
analogs of the Hermitian boundary conditions for quantum graphs
which have been derived first in the Ref.\ci{Kost}. Such
conditions are needed to construct PT-symmetric quantum graphs.
Also, we consider special cases of the  boundary conditions which
are consistent with the general ones and obtain secular equation
for finding the eigenvalues of the Schrodinger operator on graphs.
By solving numerically such secular equation we show that the
eigenvalues of the problem are real, the norm is positively
defined and the Kirchhoff rule is broken at the branching point of
a graph. Motivation for the study of PT-symmetric quantum graphs
mainly comes from the possibility for their experimental
realization in optical waveguide networks. Such networks can be
constructed by connecting optical waveguides via dissipative,
optically absorbing material. Also, condensed matter realizations
using branched graphene
nanoribbons or branched polymers can be considered.\\
{\it PT-symmetric boundary conditions.} Let us first recall construction of
Hermitian boundary conditions for quantum graphs which were derived in
\ci{Kost}. The Schr\"odinger equation on metric star graph with $N$ finite
bonds, $b_j\sim(0;L_j),\,j=1,2,...,N$ can be written as (in units
$\hbar=2m=1$)\\
\begin{equation}
-\frac{d^2\psi_j}{dx^2} =k^2\psi_j\,\,,\,x\in(0;L_j),j=1,2,3...,
N,\label{eq9}
\end{equation}
where $L_j$ length of the $j$th bond.

The inner product of two functions, $\phi$ and $\psi$ on a graph can be written
as \ci{Kost}
\begin{align}
\langle\phi,\psi\rangle=\sum_{j=1}^N\int_{0}^{L}\phi_{j}(x)\psi^*_{j}(x)dx.
\lab{inner1}
\end{align}
Here we introduce so-called skew-Hermitian product on graph, which
is defined for arbitrary differential operator, $H$ as
\ci{Kost}\be \Omega(\psi,\phi)=\langle
H\psi,\phi\rangle-\langle\psi,H\phi\rangle .\lab{skew1} \ee Then
for Eq.\re{skew1} general Hermitian boundary conditions on metric
star graph can be written as \ci{Kost}
\begin{align}
&\Omega(\psi,\phi) = -\sum_j^{N} [\phi_j^*(0)\frac{d\psi_j(0)}{dx}
-\psi_j(0)\frac{d\phi_j^*(0)}{dx}] \nonumber \\
&+\sum_j^{N} [\phi_j^*(L)\frac{d\psi_j(L)}{dx}
-\psi_j(L)\frac{d\phi_j^*(L)}{dx}]=0 \lab{skew2}
\end{align}

Eq.\re{skew2} is provides the  boundary conditions keeping the
Schr\"odinger operator  on metric star graph as self-adjoint and
can be rewritten in compact form as \ci{Kost}
\begin{equation}
{\bf A}\Psi+{\bf B}\Psi'=0,\label{skew3}
\end{equation}
where
\begin{eqnarray}
&rank({\bf A})=rank({\bf B})=2N,\nonumber\\
&{\bf AB^\dagger}={\bf BA^\dagger}.\label{skew4}
\end{eqnarray}

Our purpose is to derive PT-symmetric analogs of Eqs.\re{skew1}
and \re{skew2}. To do this, one should use  in Eq.\re{inner1}
PT-symmetric inner product, instead the Hermitian inner product.
Such inner product is given by \ci{CMB5,CMB9,CMB7}
\begin{equation}
\langle\psi,\varphi\rangle_{PT}=\underset{j=1}{\overset{N}{\sum}}\underset{0}{\overset{L_j}{\int}}PT\psi_j(x)\cdot\varphi_j(x)dx.\label{eq1}
\end{equation}
Then using the relations
\begin{equation}
P\psi_j(x)=\psi_j(L_j-x),\,T\psi_j(x)=\psi_j^*(x),\label{eq2}
\end{equation}
for Eq.\re{eq9}, we can write PT-symmetric form as
\begin{align}
\Omega(\psi,\varphi)=&\langle H\psi,\varphi\rangle-\langle\psi,H\varphi\rangle
\nonumber
\\
=&\underset{j=1}{\overset{N}{\sum}}[-\frac{\partial\psi_j^*(0)}{\partial
x}\cdot\varphi_j(L_j)+\frac{\partial\psi_j^*(L_j)}{\partial
x}\cdot\varphi_j(0)
\nonumber \\
&+\psi_j^*(0)\cdot\frac{\partial\varphi_j^*(L_j)}{\partial x}
-\psi_j^*(L_j)\cdot\frac{\partial\varphi_j^*(0)}{\partial x}].\label{eq4}
\end{align}

Furthermore, introducing the notations
\begin{align}
\Psi=(\psi_1(0),\psi_2(0),&...,\psi_N(0),\nonumber \\ &\psi_1(L_1),\psi_2(L_2),...,\psi_N(L_N))^T,\nonumber\\
\Psi'=(-\psi_1'(L_1),-&\psi_2'(L_2),...,-\psi_N'(L_N),\nonumber
\\ &\psi_1'(0),\psi_2'(0),...,\psi_N'(0))^T.\label{eq5}
\end{align}

PT-symmetric product for Eq.\re{eq9} can be rewritten as
\begin{equation}
\Omega(\psi,\varphi)=\left(\Phi^T\,\Phi'^T\right)\left(\begin{array}{cc}0&I_{2N}\\-I_{2N}&0\end{array}\right)\left(\begin{array}{cc}\Psi^*\\\Psi'^*\end{array}\right).\label{eq6}
\end{equation}
It is clear that  $\Omega(\psi,\varphi)=0$, if Eq.\re{skew4} holds
true. Although the general boundary conditions given by
Eq,\re{eq6} provide PT-symmetry of the Schrodinger operator on
graph and reel eigenvalues, the norm of the wave function of such
system is not positively defined \ci{CMB5}. However, physically
correct PT-symmetric boundary conditions providing both, real
eigenvalues and positively defined norm can be obtained using an
extension of the above PT-symmetric inner product, which is called
CPT-symmetric inner product. It was derived first in \ci{CMB5} and
studied in detail later in \ci{CMB9,CMB11}. With such
CPT-symmetric inner product determined by \ci{CMB5,CMB9,CMB11}
\begin{equation}
\langle\psi,\varphi\rangle_{CPT}=\underset{j=1}{\overset{N}{\sum}}\underset{0}{\overset{L_j}{\int}}CPT\psi_j(x)\cdot\varphi_j(x)dx.\label{eq1}
\end{equation}
where
$$
C_j(x,y)=\phi_j^{(n)}(x)\phi_j^{(n)}(y),\label{eq4}
$$ and using
$$
CPT\psi_j(x)=\underset{0}{\overset{L_j}{\int}}C(x,y)\psi_j^*(L_j-y)dy,\label{eq3}\\
$$
we can write  CPT-symmetric boundary conditions for Eq.\re{eq9} as
\begin{align}
\Omega(&\psi,\varphi)=2\underset{j=1}{\overset{N}{\sum}}\left[-\varphi_j(L_j)\cdot\frac{\partial}{\partial
y}\psi_j^*(0)+\frac{\partial}{\partial
y}\varphi_j(L_j)\cdot\psi_j^*(0)\right.\nonumber
\\ &\left.+\varphi_j(0)\cdot\frac{\partial}{\partial
y}\psi_j^*(L_j)-\frac{\partial}{\partial
y}\varphi_j(0)\cdot\psi_j^*(L_j)\right] = 0.\label{eq006}
\end{align}
In derivation of eq.\re{eq006} we took into account that the
eigenvalues of Eq.\re{eq9} are real. It is clear that
Eq.\re{eq006} can be written in compact form given by
Eq.\re{skew4}. \\
{\it Secular equation.} Let us now obtain explicit solutions of
Eq.\re{eq9} for some PT-symmetric boundary conditions. General
(without boundary conditions) solution of Eq.\re{eq9}can be
written as
\begin{equation}
\psi_j(x)=A_j\cos{k(L_j-x)}+B_j\sin{k(L_j-x)}.\label{eq11}
\end{equation}
The eigenvalues, $k_n$ and constants $A_j,$ $B_j$ can be found
from the boundary conditions. A set of boundary conditions which
are consistent with Eq.\re{eq006} can be written as (here, for
simplicity we consider metric star graph with three bonds)
\begin{eqnarray}
&\psi_1(0)=\psi_2(0)=\psi_3(0),\nonumber\\
&\dfrac{\partial\psi_1(L_1)}{\partial
x}+\dfrac{\partial\psi_2(L_2)}{\partial
x}+\dfrac{\partial\psi_3(L_3)}{\partial x}=0,\nonumber\\
&\psi_j(L_j)=0,\,j=1,2,3.\label{bc01}
\end{eqnarray}
Such boundary conditions have been derived in \ci{Znojil}. Another
set of boundary conditions which is consistent with the general
ones given by Eqs. \re{eq006} is
\begin{eqnarray} &\left.\frac{\partial\psi_1}{\partial
x}\right|_{x=0}=\left.\frac{\partial\psi_2}{\partial
x}\right|_{x=0}=\left.\frac{\partial\psi_3}{\partial
x}\right|_{x=0},\nonumber\\
&\psi_1(L_1)+\psi_2(L_2)+\psi_3(L_3)=0,\nonumber\\
&\left.\frac{\partial\psi_j}{\partial
x}\right|_{x=L_j}=0,\,\,\,j=1,2,3.\label{bc001}
\end{eqnarray}
Both sets of boundary conditions lead to the same secular equation
which is given by \be \sin kL_1\sin kL_2+\sin kL_1\sin kL_3+\sin
kL_2\sin kL_3=0. \lab{secular1}\ee The eigenfunctions
corresponding to boundary conditions Eq.\re{bc01} can be written
as
\begin{equation}
\psi_j(x,k_n)=B_n\frac{\sin k_n(L_j-x)}{\sin
k_nL_j},\label{eigfun1}
\end{equation}
while for the boundary conditions \re{bc001} we have the
eigenfunctions \be \psi_j(x,k_n)=A_n\frac{\cos k_n(L_j-x)}{\sin
k_nL_j},\label{eigfun2}]\ee where $A_n$ and $B_n$ are the
normalization constants which are given by
$$
B_n=\left(\underset{j=1}{\overset{3}{\sum}}\frac{2k_nL_j-\sin2k_nL_j}{4k_n\sin^2k_nL_j}\right)^{-\dfrac12}
$$
and
$$
A_n=\left(\underset{j=1}{\overset{3}{\sum}}\frac{2k_nL_j+\sin2k_nL_j}{4k_n\sin^2k_nL_j}\right)^{-\dfrac12}.
$$
\begin{figure}[th!]
\includegraphics[width=92mm]{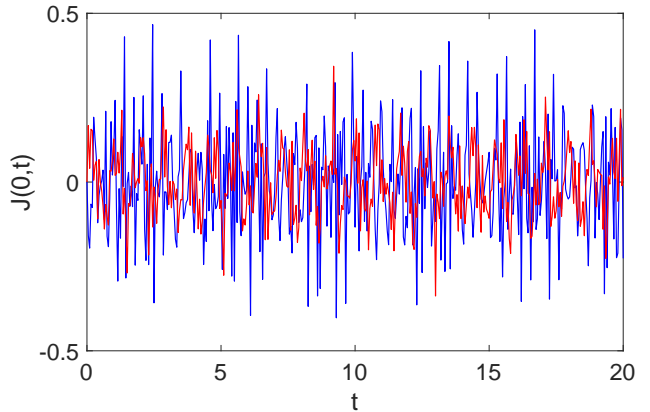}
\caption{(Color online) Total current at the vertex for the
boundary conditions \re{bc01} (blue) and \re{bc001} (red) as a
function of time (for ).} \label{pic1}
\end{figure}

It is clear that the norms of the eigenfunctions given by
Eqs.\re{eigfun1} and \re{eigfun2} are positively defined.  Direct
numerical computation of the roots of Eq.\re{secular1} shows that
they are indeed  real. This implies that the eigenvalues of
Eq.\re{eq9} for the boundary conditions \re{bc01} and \re{bc001}
are real. Unlike Hermitian quantum graphs (see, e.g.
Refs.\ci{Kost,Uzy1,Uzy2,Exner15}), the boundary conditions given
by Eqs.\re{bc01} and \re{bc001} do not provide Kirchhoff rules.
This implies breaking of the current conservation at the branching
point that can be directly checked by  computing numerically total
current at the vertex ($x=0$) given by
$$
J(0,t) =J_1(0,t) +J_2(0,t) +J_3(0,t),
$$
where
$$
J_j(0,t)=\frac{i}{2}[\psi_j(0,t)\partial_x\psi_j^*(0,t)-\partial_x\psi_j(0,t)\psi_j^*(0,t)],
$$
is the current on each bond and
$$
\psi_j(x,t)=\underset{n}{\sum}C_ne^{-ik_n^2t}\phi_j(x,k_n).
$$

In Fig.~\ref{pic1} the total current at the vertex, $J(0,t)$ for
the boundary conditions \re{bc01} and \re{bc001} are plotted as a
functions of time. It is
clear that current conservation (Kirchhoff rule) is broken.\\
{\it Experimental realization on branched optical waveguides.}
Some models for PT-symmetric networks have been discussed earlier
in the literature \ci{PTSN1,PTSN2}. However, the above model of
the PT-symmetric quantum graph can be easily realized using
branched (Y-junction) optical waveguides which are connected
according to the boundary conditions given by Eqs. \re{bc01} or
\re{bc001}. It is clear that both set of boundary conditions
provide absence of current at the end of the branches and the
continuity of the wave function at the branching point. Taking
into account breaking of Kirchhoff rules in these boundary
conditions, one may construct such PT-symmetric quantum graph by
connecting three optical wave guides via the small-size partially
absorbing optical material. The  ends of the branches of of the
waveguides should provide total reflection of the wave. Similarly,
one may consider experimental realization in general (more than
three branched) star graph of optical waveguides and arbitrary
graph topology such us, e.g. tree, loop and complete graphs. In
this case all the branching points should be optically absorbing
material, while the edge branches should provide zero-current at
the ends. One of the options for dissipatively coupled optical
waveguides have been recently discussed in \ci{Mukherjee}.
Different branched versions of such system can be also good
candidate for PT-symmetric quantum graph. \\
{\it Conclusions.} We have studied the problem called
``PT-symmetric quantum graphs'' which represents branched quantum
wires, whose branches are  connected according to PT-symmetric
rules. The latter implies that the boundary condition at the
branching points and ends of the bonds provide PT-symmetry of the
Schr\"odinger operator on a graph. General boundary conditions
providing PT- symmetry of the Schr\"odinger operator on a graph
and having real eigenvalues, as well as positively defined norm of
the eigenfunctions are derived. Special types of the boundary
conditions following from such conditions are presented. It is
shown that the eigenvalue spectrum of the Schr\"odinger equation
on metric graphs for such boundary conditions is real.
Experimental realization of PT-symmetric quantum graph using
branched optical waveguides is discussed. The above approach used
for constructing PT-symmetric quantum star graph is applicable for
other graph topologies, such us tree graph, loop graph, different
fractal graphs, etc. Also, extension to the relativistic case  can
be easily done for PT-symmetric
 Dirac and Klein-Gordon equations on metric graphs.\\
{\it Acknowledgements.} We thank Carl M. Bender for his valuable
comments on the paper. This work is partially supported by a grant
of the Ministry of Innovation Development of Uzbekistan (Ref. No.
 F-2-003).

\end{document}